# PICO - the probe of inflation and cosmic origins

Brian M. Sutin[*,a], Marcelo Alvarez[b], Nicholas Battaglia[w], Jamie Bock[d], Matteo Bonato[e], Julian Borrill[f], David T. Chuss[g], Joelle Cooperrider[a], Brendan Crill[a], Jacques Delabrouille[h], Mark Devlin[i], Thomas Essinger-Hileman[r], Laura Fissel[j], Raphael Flauger[k], Krzysztof Gorski[a], Daniel Green[m], Shaul Hanany[n], Johannes Hubmayr[o], Bradley Johnson[p], William C. Jones[c], Lloyd Knox[q], Alan Kogut[r], Charles Lawrence[a], Jeff McMahon[s], Tomotake Matsumura[t], Mattia Negrello[u], Roger O'Brient[a], Christopher Paine[a], Clement Pryke[n], Peter Shirron[r], Amy Trangsrud[a], Qi Wen[n], Karl Young[n], Gianfranco de Zotti[v]

[a]Jet Propulsion Laboratory, California Institute of Technology, USA; [b]University of California Berkeley, USA; [c]Princeton University, USA; [d]California Institute of Technology, USA; [e]Osservatorio di Radioastronomia, Italy; [f]Lawrence Berkeley National Laboratory, USA; [g]Villanova University, USA; [h]Laboratoire AstroParticule et Cosmologie and CEA/DAP, France; [i]University of Pennsylvania, USA; [j]NRAO, USA; [k]University of California San Diego, USA; [m]University of Toronto, Canada; [n]University of Minnesota, Twin Cities, USA; [o]NIST, USA; [p]Columbia University, USA; [q]University of California Davis, USA; [r]Goddard Space Flight Center, USA; [s]University of Michigan, USA; [t]University of Tokyo, Japan; [u]Cardiff University, UK; [v]Osservatorio Astronomico di Padova, Italy; [w]Center for Computational Astrophysics, Flatiron Institute, USA

## ABSTRACT

The Probe of Inflation and Cosmic Origins (PICO) is a NASA-funded study of a Probe-class mission concept. The top-level science objectives are to probe the physics of the Big Bang by measuring or constraining the energy scale of inflation, probe fundamental physics by measuring the number of light particles in the Universe and the sum of neutrino masses, to measure the reionization history of the Universe, and to understand the mechanisms driving the cosmic star formation history, and the physics of the galactic magnetic field. PICO would have multiple frequency bands between 21 and 799 GHz, and would survey the entire sky, producing maps of the polarization of the cosmic microwave background radiation, of galactic dust, of synchrotron radiation, and of various populations of point sources. Several instrument configurations, optical systems, cooling architectures, and detector and readout technologies have been and continue to be considered in the development of the mission concept. We will present a snapshot of the baseline mission concept currently under development.

## 1. INTRODUCTION

### 1.1 NASA's "Probe" class concept

In astronomy and astrophysics, NASA currently flies small and medium Explorer missions (<$250M), as well as multi-billion-dollar flagship observatories like JWST and WFIRST. There are a number of science opportunities that are beyond the scope of the Explorer program, but that don't require flagship-level funding. To explore these opportunities, NASA has funded studies of 10 "Probe" class ($400M-$1B) mission concepts. The Probe of Inflation and Cosmic Origins (PICO) is one of these mission studies. Reports of these mission studies are due to NASA at the end of the 2018 and NASA's plan is to forward the reports to consideration by the next Astronomy and Astrophysics Decadal panel. This

---

[*] brian.m.sutin@jpl.nasa.gov; phone 1 818 354-2227; https://www.jpl.nasa.gov

paper comes part way through the PICO study, and describes a snapshot of the instrument design at this time of the study.

## 1.2 PICO Science

Electromagnetic radiation in the millimeter and sub-millimeter wavelength band contains a wealth of information about the origin, structure, and evolution of the Universe. With 21 frequency bands centered between 21 and 799 GHz and polarization sensitivity across this spectrum, PICO has unprecedented sensitivity to information encoded in the cosmic microwave background (CMB) radiation and in Galactic emissions. Measurements of these sources would reveal new information about the structure and dynamics of the Milky Way, detect the signature (or constrain models of) inflation, measure the properties of the fundamentals particles of nature, and provide key information on the evolution of Galactic and extragalactic structures across cosmic time.

Quantum fluctuations of the space-time metric at times as short as $10^{-35}$ seconds after the big bang are predicted to generate gravitational waves that imprint a signature on the polarization of the CMB. A detection of this signature sets the energy scale at which an inflationary period occurred near the big bang, and thus provides constraints on the physics of inflation. These constraints are not attainable in other ways. PICO would detect the energy scale of inflation or place an upper limit that would exclude broad classes of potentials as the driving force for inflation.

The first stars in the Universe formed within about 1 billion years after the big bang. As they formed they ionized the surrounding medium, which was dominated by neutral hydrogen. Many details of this process of 'reionization' are not known. PICO would determine the history of star formation in the Universe through its measurement of the optical depth to reionization $\tau$. The measurement requires scanning large portions of the sky, a requirement particularly well suited for space missions. PICOs measurement uncertainty on $\tau$ would be limited only by cosmic statistics, not instrument noise.

Structure formation in the Universe depends on the masses of neutrinos. Lensing of CMB photons as they traverse the Universe reveals the properties of cosmic structures, and is thus a sensitive probe of the masses of neutrinos, specifically the sum of neutrino masses $\Sigma m_\nu$. Extracting $\Sigma m_\nu$ from the lensing measurement also requires knowledge of $\tau$ and of the matter density in the Universe. PICO would probe $\Sigma m_\nu$ by mapping the effects of lensing with signal-to-noise ratio of several hundred and combining the measurement with its own high precision measurement of $\tau$ with the measurement of the matter density forthcoming from already funded cosmic surveys.

In addition to the three known neutrino families, the Universe may harbor other similar light relics. The number of such relic species is quantified through the parameter $N_{eff}$, for which the current measured value[2] is $3.04 \pm 0.18$. The theoretical value in the presence of only three neutrino families is 3.046, and in the standard model of particle physics the smallest increment, if other relic species existed in early Universe, is 0.027. PICO would probe for additional relic species through their effect on the spatial anisotropy of the polarization of the CMB.

Star formation is regulated by magnetic fields, turbulence, and feedback from young stars, but the relative importance of these mechanisms is not well understood. Detailed mapping of the magnetic field structure on a broad range of scales would transform our understanding of star formation and of the role of magnetic fields in Galactic evolution. Elongated Galactic dust grains align on average with respect to the local magnetic fields and emit polarized radiation at sub-mm to mm wavelengths. PICO would map the structure of magnetic fields by detecting this polarized emission with resolutions and depths not previously attained.

PICO would produce a complete census of cold dust in the nearby universe as well as a rich catalog of newly discovered extra-galactic sources distributed over the entire sky. These sources include highly magnified dusty galaxies, and proto-clusters detected via the sub-mm emission of their member galaxies. PICO would utilize the thermal Sunyaev-Zel'dovich effect (SZE) to produce a catalog of galaxy clusters. PICO would also characterize the polarization properties of many radio and FIR emitting galaxies. This rich catalog would be used to probe star formation history, determine galaxy and cluster formation and evolution, learn about the properties of dark matter, and study radio jets in radio loud sources.

## 2. INSTRUMENT DESIGN CONTEXT

### 2.1 Instrument Requirements

PICO requires imaging polarimetry of the CMB to search for inflationary gravitational waves, constrain the sum of the neutrino masses, measure the optical depth to reionization and detect clusters and proto-clusters through the SZE.

Mapping the CMB requires strong sensitivity at frequencies between 30 and 300 GHz. Frequencies up to 600 GHz are required to generate a large census of SZ sources. Broad frequency coverage at lower and higher frequencies is required to identify and separate Galactic sources of emission by identifying their spectral signatures. Extending to even higher frequency bands improves source separation, and is required to provide the information about the role of Galactic magnetic fields. Intensity and polarization measurements across the spectrum give information about cosmic structure formation and evolution.

The inflation and reionization science goals require measurements of correlated polarization on the largest angular scales available across the sky, so PICO requires a full sky survey. Diffraction limited resolution matching a ~1.4 m aperture telescope is required to achieve the entire array of science goals.

PICO's ability to achieve its science goals is strongly driven by the sensitivity of the sky maps that it produces. This, in turn, is a function of the individual detector sensitivity, the number of detectors on the focal plane, and the duration of the mission. PICO instrument sensitivity is discussed by Young et al[3]. To achieve a sensitivity limited only by the photon noise from incident astrophysical signals, PICO uses transition edges sensor (TES) bolometers that are cooled to 0.1 K. Because the PICO instrument observes in the mm and sub-mm wavelength band, thermal emission by the instrument itself can introduce noise in the detectors. This sets thermal requirements on the larger instrument structures, including the reflectors and the aperture stop.

Because the Probe concepts are under consideration for the next decade, they are forward looking and can recommend technologies that do not yet meet the criteria for NASA's Technology Readiness Level (TRL) 5, but that could be developed to that level by ~2023.

NASA requires that Probes meet the requirements associated with a NASA Class B risk classification. This classification is based on factors such as the national significance and cost of the mission, and it defines the expectations for risk mitigation. For Class B missions, essential spacecraft and instrument functions are typically fully redundant.

NASA requires that Probes be compatible with an Evolved Expendable Launch Vehicle (EELV). The launch vehicle defines the mass-to-orbit capability, the launch environments (e.g. vibrations), and the fairing volume available for the Observatory. In the development of the PICO mission concept, the Falcon 9 is used as the reference launch vehicle. The Falcon 9 launch vehicle offers a 5.2 m (outer diameter) fairing with a 4.6 m diameter payload dynamic envelope.

## 2.2 Concept of Operations

The PICO concept of operations is similar to that of the successful WMAP and Planck missions. After launch, PICO cruises to an orbit around the Earth-Sun L2 Lagrange point. A decontamination period is followed by instrument cooldown. After in-orbit checkout is complete, PICO begins its science survey. PICO has a single science observing mode, surveying the sky using a pre-planned repetitive survey pattern. Instrument data are compressed and stored on-board, then returned to Earth in daily Ka-band science downlink passes (concurrent with science observations). Because PICO is observing relatively static galactic, extragalactic, and cosmological targets, there are no requirements for time-critical observations or data latency. Presently, there are no plans for targets of opportunity or guest observer programs during the prime mission.

To scan the sky, PICO spins with a period of $T_{spin}$ = 1 minute about a spin axis oriented at angle $\alpha$ = 26° from the anti-solar direction. The telescope boresight is oriented at an angle $\beta$ = 69° away from the spin axis. This $\beta$ angle is chosen to give $\alpha + \beta$ = 95°, enabling full sky mapping. The spin axis is forced to precess about the anti-solar direction with a period $T_{prec}$. Optimization of $T_{prec}$ is ongoing; we are considering periods between 10 and 48 hours. The angular separation of sequentially mapped rings is therefore between 16 and 3 arcminutes. The anti-solar direction tracks with the Earth in its yearly orbit around the Sun, so this scan strategy maps the full sky every 6 months.

A robust single-fault-tolerant attitude control system spins the PICO instrument utilizing an architecture with heritage from the system that spins the SMAP mission's 6-m instrument antenna at 14.6 rpm. For PICO, spinning at 1 rpm (like the Planck mission), the $l$=2 quadrupole power spectral mode would appear in the data timestream[4] at $2l \sin(\beta)/T_{spin}$ = 0.06 Hz. This enables relaxed requirements on 1/f noise relative to slower spin rates. Sampling the detectors at a rate that corresponds to 1/3 of a beam width at 1 rpm, and assuming 4x compression, PICO can downlink its science data to the Deep Space Network in 4hr daily passes (a reasonable time allocation for a Probe-class mission).

To measure polarization, PICO differences the signals from pairs of detectors that have polarization sensitivity that differs by 90°. Mismatch in the antenna response of the two detectors can introduce systematic errors in the polarization map. Selecting an observation strategy that scans across the same sky pixel with a diversity of orientations reduces this effect. PICO's α angle (26°) was chosen to be substantially larger than the Planck mission's α angle (7.5°) to provide increased mitigation of systematic effects associated with differences between the gain calibrations of detector pairs and differences in the antenna response of detector pairs[26].

Like Planck, the PICO orbit around L2 is designed to be small enough to ensure than the Sun-Probe-Earth (SPE) angle is less than 15° while on orbit. This keeps the telescope boresight at least 70° away from the Earth at all times during the survey, mitigating the impacts of terrestrial and lunar stray light.

## 3. INSTRUMENT

### 3.1 Instrument Architecture

A block diagram of the instrument is shown in Figure 1. During the survey, the instrument is spun at 1 rpm. Spacecraft control is simplified by mounting the instrument and coolers on a spinning platform, while other spacecraft elements, including the inertia wheels, are housed on a stationary module.

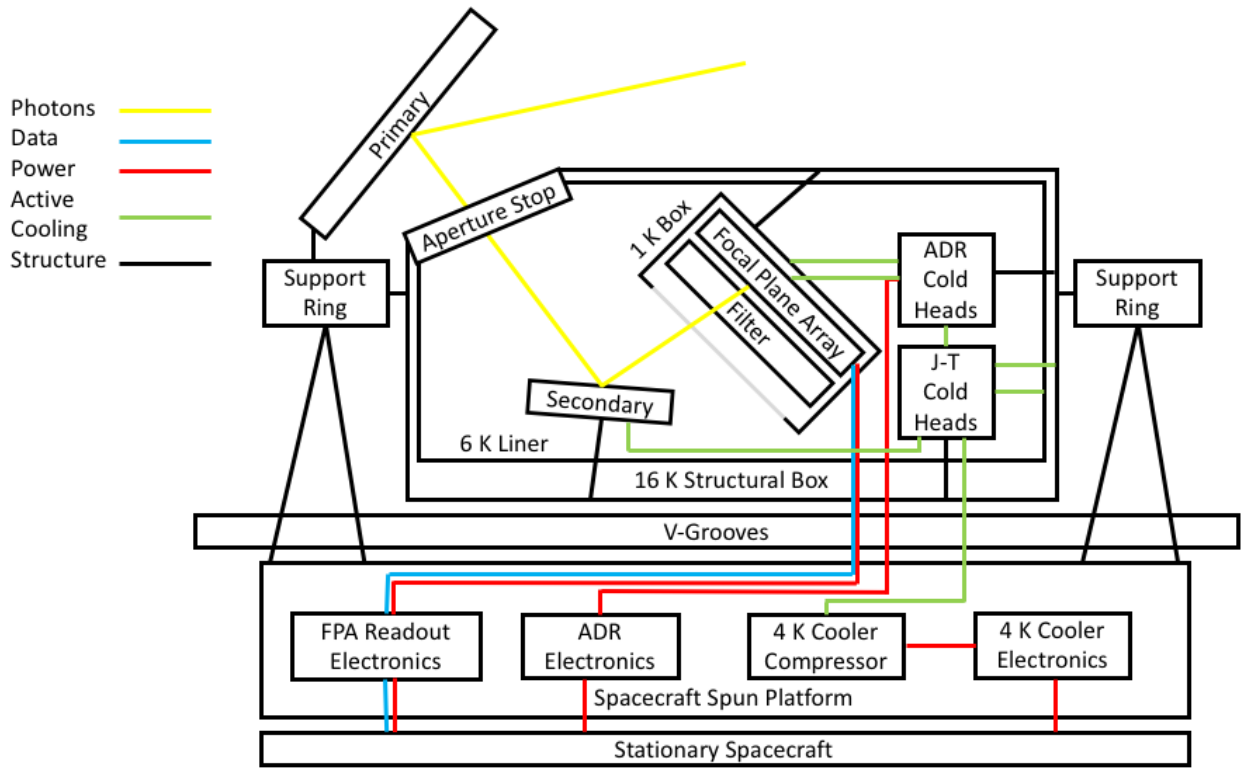

Figure 1. PICO system block diagram.

The instrument is built around a Dragone-style two-reflector telescope with an internal aperture stop between the primary and secondary. A structural ring, supported off of the spacecraft by thermal isolating bipods, supports the primary and a structural box containing and supporting the rest of the instrument optics, focal plane, and thermal heads. A set of V-groove radiators, mounted off of these bipods, keeps sunlight off of the telescope and thermally isolates the instrument from the spacecraft. The Joule-Thompson (J-T) cooler head maintains the structural box at 16 K, as well as supplying 4 K cooling to the adiabatic demagnetization refrigerator (ADR), the secondary reflector, the aperture stop, and a 6 K liner inside the 16 K structural box. The properties of the J-T provide the coldest temperature, either 16 or 4 K, at a single point. Other points along the return path of the gas would be warmer. The ADR cools the focal plane to 0.1 K and also provides heat sink point at 1 K. Heat sources such as the FPA readout electronics, ADR cooler electronics, 4 K cooler compressor, and 4 K cooler electronics are all placed on the spinning portion of the spacecraft.

### 3.2 Optical design

Young et al.[3] describes the optical system - a two-reflector telescope with a 1.4m effective aperture - and presents the design of the focal plane.

### 3.3 Detectors and Electronics

#### 3.3.1 Detector Technology

PICO's focal plane would use an imaging array of TES bolometers to map the CMB and its galactic foregrounds in 25% wide bands with band centers ranging from 21-799 GHz. All but the highest frequency channels would feed three photometric channels with a common broad-band antenna. Current designs have small spectral gaps between the filters' passbands where the reactance strongly reflects power, although alternative log-periodic designs have been prototyped that would guarantee contiguous channels with high optical efficiency[5]. Several competing technologies have matured over the past ten years using horn-coupling[6], antenna-array coupling[7], and sinuous antenna/lenslet-coupling[8], delivering high optical efficiency over more than an octave of bandwidth. Microstrip mediates the signals between the antenna and detectors in all of these schemes and partitions the feed's wide continuous bandwidth into narrow channels with integrated micro-machined filter circuits[5]. Suborbital experiment teams have implemented these technologies for frequency bands that are relevant for observations from the ground[6,7,9]. There are proposed balloon missions to extend the technologies to higher frequencies[10].

PICO's highest frequency channels are beyond the Niobium superconducting band-gap, rendering microstrip filters a poor solution for defining the optical pass band. Instead, we would couple those detectors through horns directly to the absorber coupled bolometers in the throat of a waveguide and rely on available, quasi-optical metal-mesh filters to define an upper band edge opposite the lower edge from the waveguide cut-off. Numerous experiments[11,12] successfully used this approach, including the Planck satellite[13].

Myriad sub-orbital submillimeter experiments have demonstrated TESs covering channels from 30-410 GHz, showing white noise with stability to at least as low as 20 mHz[14]. Several laboratory experiments show that these detectors can be made background limited in the low loading environment they would experience at L2 and that they are stable well below the planned rotation rate of the spacecraft[14]. A conceptual layout of the focal surface is shown in Figure 2. For a detailed report on the focal plane design, see Young et al.[3] in this volume.

#### 3.3.2 Readout Electronics

Suborbital experiment teams over the past ten years have chosen to uses voltage biased TESs because their current readout lends itself to SQUID-based multiplexing. Multiplexing reduces the number of wires to the cryogenic stages and thus the total thermal load that the cryocoolers must dissipate. This approach also simplifies the instrument design. In the multiplexing circuitry, SQUIDs function as low noise amplifiers and cryogenic switches.

Two common and mature multiplexing schemes have emerged. One is a frequency division multiplexer (FDM) that AC biases each detector in a set with a unique ~100 kHz tone, sums the channels on a common line, and then amplifies in a SQUID series array (SSA) that is typically placed at the 4 K stage[15]. The other mature alternative is a time division multiplexer (TDM) that assigns the detectors address in a square matrix of commonly read columns and sequentially cycles through each row of the array[16].

In contrast to FDM, TDM uses more complex cryogenic circuitry with numerous SQUIDs per column and higher thermal loading through a higher wire-count. However, it has smaller ambient temperature electronics power consumption because the circuits only provide DC biases, feedback, and digitization at ~100 kHz. In FDM ambient temperature electronics are required to synthesize and readout all the tones. The thermal loading from the wire harnesses, while higher than FDM cables, is still subdominant to radiative loading in PICO.

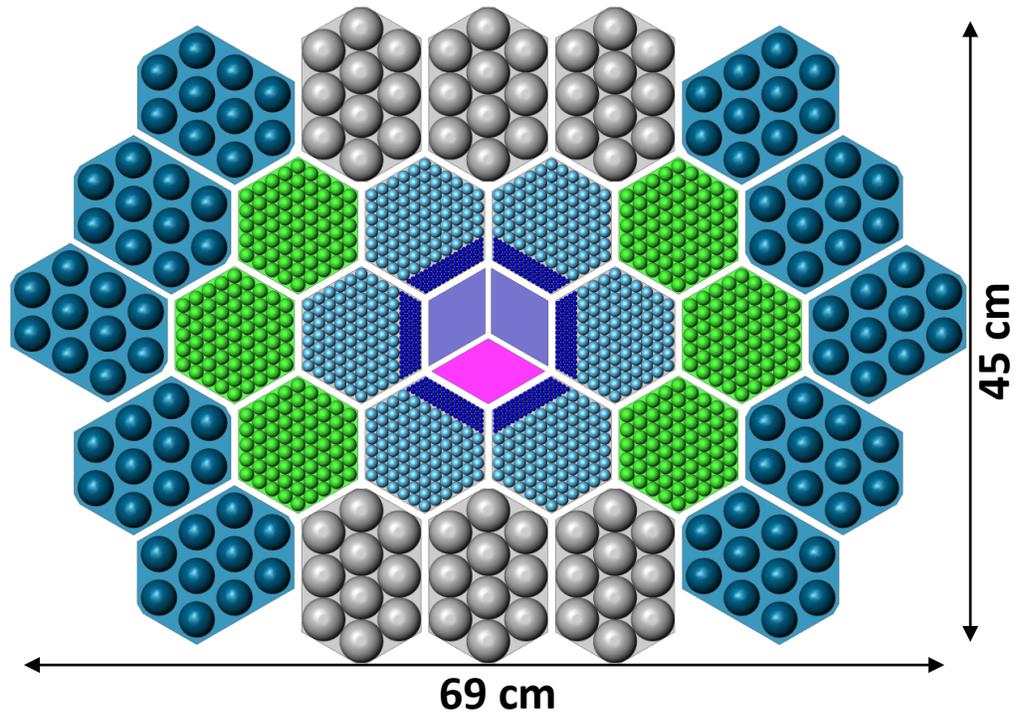

Figure 2. PICO focal surface layout. Detectors are fabricated on 6 types of silicon wafers (colored tiles in this figure). The wafers are placed on the focal surface such that higher frequency bands, which require better optical performance, are placed nearer to the center. See Young et al.[3] for more details.

The current baseline for PICO is to use TDM with a matrix of 128 rows and 102 columns. The ambient readout electronics, realized with commercially available ASICs, would consume 50 W, fitting well within the expected power limits of the instrument. Further optimization of FDM before the next phase of PICO's study may close this gap in electronics power consumption.

Future versions of the PICO instrument, perhaps when implemented in the next decade, may choose to use even larger focal planes. There are rapidly developing technologies such as microwave frequency MUX[17], kinetic inductance detectors (KIDs)[18], and thermal KIDs (TKIDs)[19], that may more sensibly address those needs if they are sufficiently mature.

### 3.4 Mechanical design & configuration

The main structure of the instrument is a ring, supported by bipods from the spacecraft. A set of four V-groove radiators, supported by the bipods, shield the instrument from the Sun and the spacecraft. The passively-cooled primary reflector and a 16 K actively cooled box are supported directly off of the main ring. The secondary reflector, and other lower temperature elements such as ADR, aperture stop, and focal plane are all supported by the 16 K box (Figures 1 & 3).

The 4.5 m diameter of the outer V-groove radiator is designed to fit within a Falcon 9 payload fairing. The shadow cone cast by the V-groove radiators as shown in Figure 3 is 29°. This corresponds to the angle to the Sun, $\alpha = 26°$, plus a 3° margin to account for the angular radius of the sun (0.5°), spacecraft pointing error, design margin, and mechanical alignment tolerances.

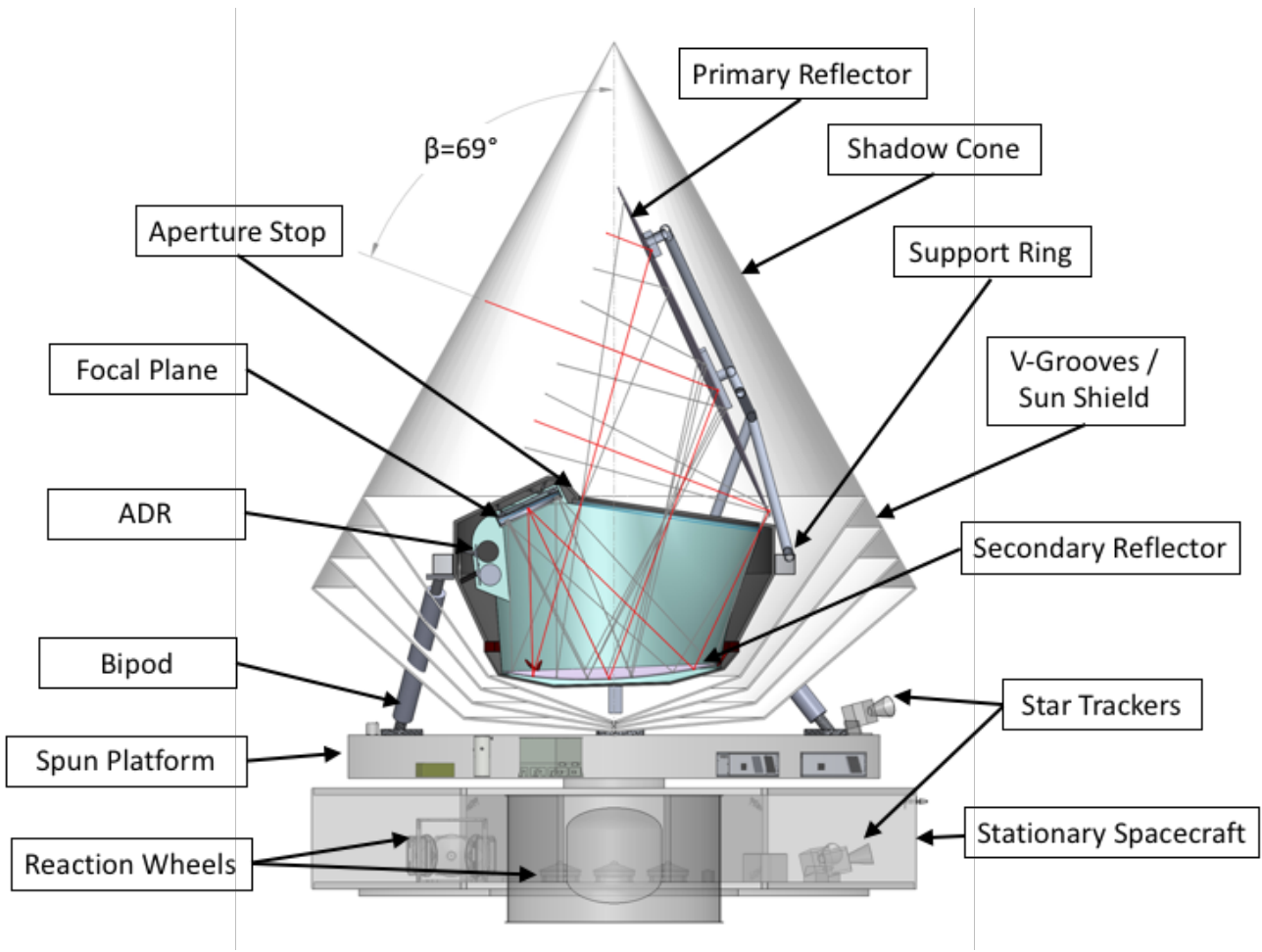

Figure 3. PICO instrument configuration.

### 3.5 Instrument Thermal Design

PICO utilizes a combination of passive and active cooling elements. The focal plane is cooled to 0.1K to support the operation of the superconducting bolometric detectors, and the reflectors, aperture stop, and other instrument elements are cooled to minimize the coupling of the detectors to radiation from the instrument itself.

#### 3.5.1 V-groove radiators

A set of four V-groove radiators give passive cooling of primary reflector and the rest of the instrument. The largest of the four V-groove shields provides a shadow cone for the inner grooves and for the instrument. The additional V-grooves radiate freely to space, each reaching successively cooler temperatures, with the final V-groove temperature near 40 K. The V-groove assembly is supported from the spacecraft bus by attachment to the low-conductance bipod struts. Similar structures are a standard approach for passive cooling, having been used on prior missions[20,21,22].

The bipod struts also carry the mechanical loads on the structural ring which supports the primary reflector (~20 K), the structural box (~16 K), the secondary reflector (at ~10 K), the thermal liner surrounding the secondary and serving as an aperture stop (~6 K), and the focal plane and sub-K adiabatic refrigerator structures. All of these components except the primary reflector are actively cooled, as discussed in the next sections.

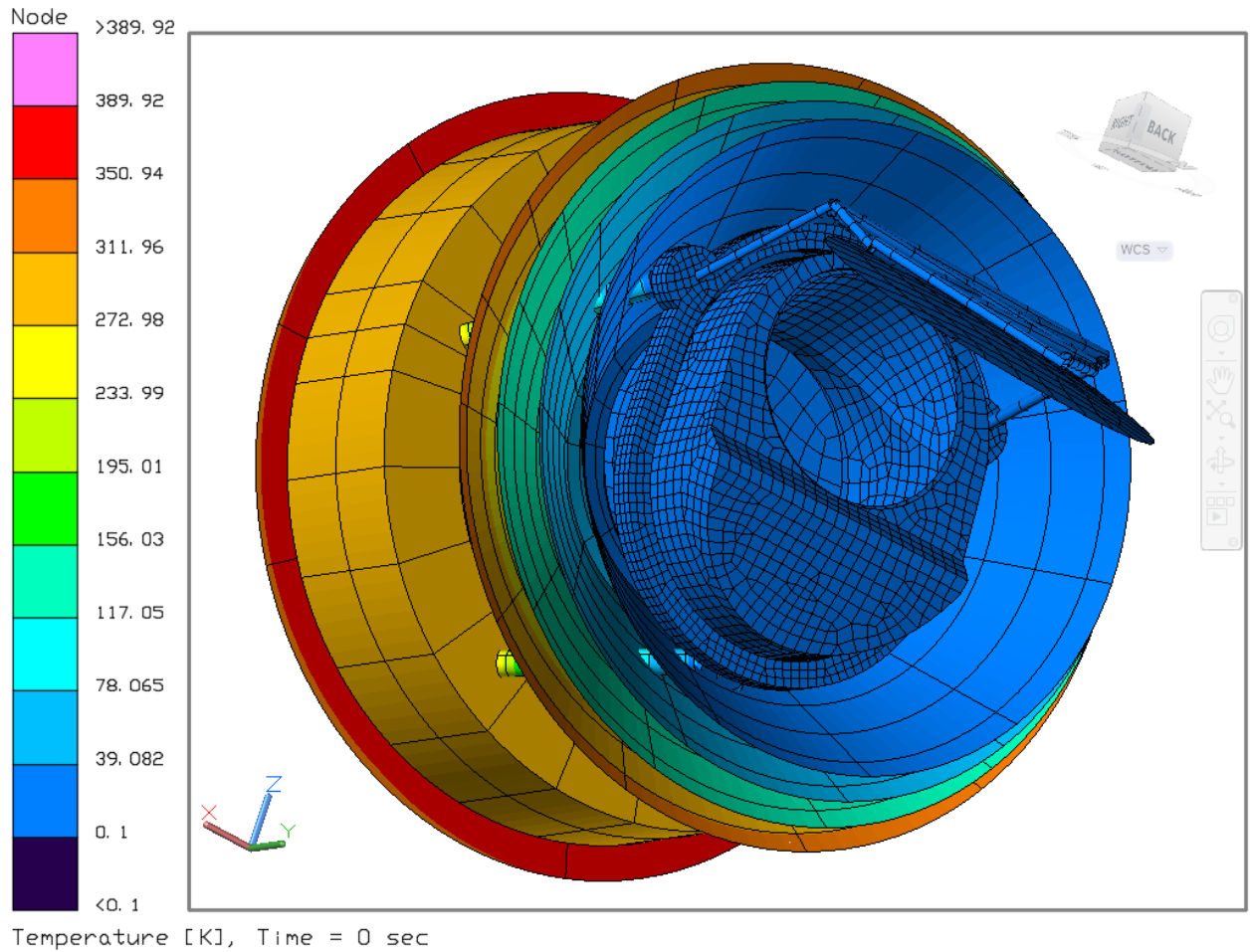

Figure 4. Thermal model of PICO temperature distribution.

### 3.5.2 ADR — sub-K cooling

A multi-stage continuous adiabatic demagnetization refrigerator (CADR) cools the PICO focal plane to 0.1 K, and the surrounding filter box structure and readout electronics to 1 K. The cADR employs 1) three stages stacked serially absorbing heat from the focal plane at 0.1 K and rejecting it to 1 K; and 2) two stages operating in parallel absorbing this rejected heat, cooling other components to 1 K, and rejecting heat at 4.5 K to the J-T stage of the mechanical cooler. This multiple parallel stage configuration provides continuous cooling with small temperature variation at both the 0.1 K and 1 K stages. Heat straps connect the two cADR cold stages to multiple points on the focal plane assembly, which itself has high thermal conductance paths built into it, to provide spatial temperature uniformity below 100 µK during operation. Heat loads in the range of 20 µW at 0.1 K and 1 mW at 1 K (time-average) are within the capabilities of current CADRs developed by GSFC.

### 3.5.3 4 K cooler

A cryocooler system similar to that used on JWST to cool the MIRI detectors[24,25] removes the heat rejected from the CADR, and cools the thermal liner and secondary reflector. The cryocooler system consists of 1) a 3-stage acoustic-Stirling ("pulse tube" or PT) cryocooler producing 16 K at its coldest point, which precools the gas stream from 2) a similar compressor modified for DC flow which drives a circulated-gas-loop incorporating J-T expansion to further cool the gas to 4.5 K. The motorized compressor and the cold heads of the PT, and the J-T circulator compressor, are located on the warm spacecraft, with relatively short lengths of tubing conducting the gas flow from the 16 K precooling point to the J-T expansion point.

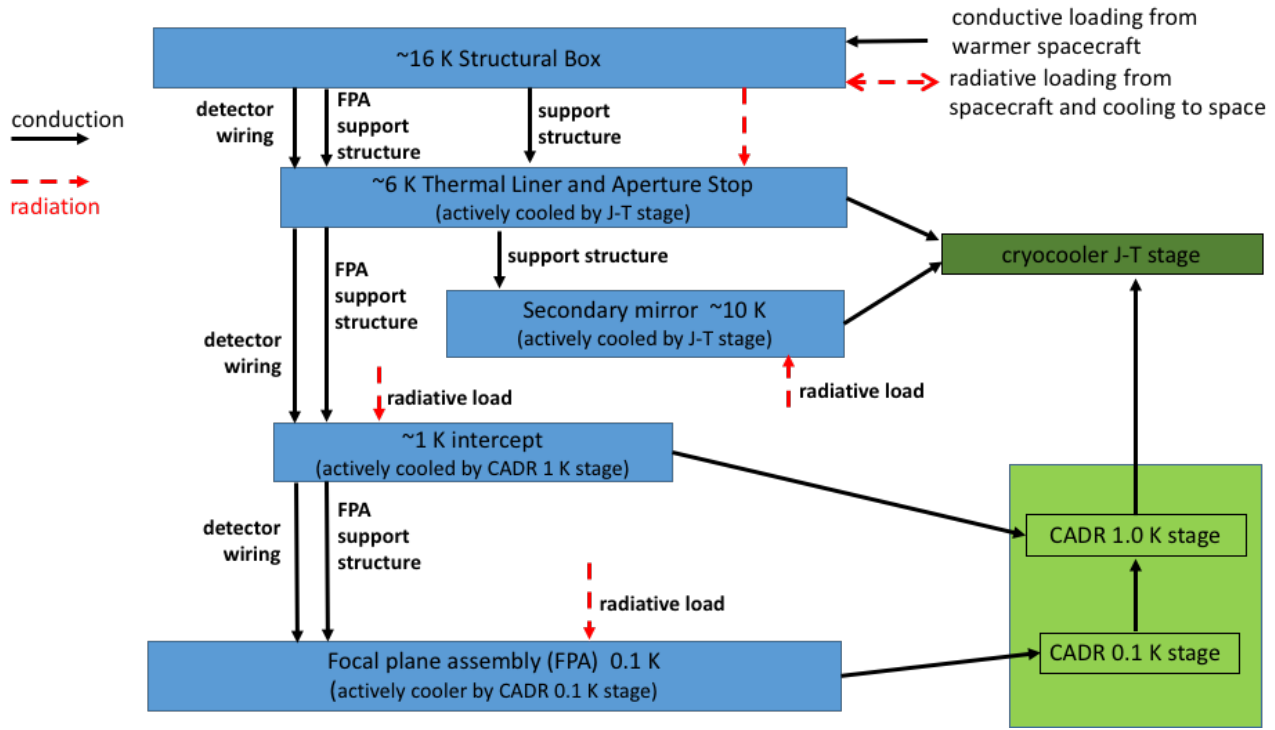

Figure 5. PICO heat flow map for the instrument assembled inside the V-groove structure.

The J-T expansion point is located at or close to the cADR heat rejection point, thereby providing the lowest temperature to the CADR. Subsequent to cooling the CADR, the returning gas flow cools the the thermal liner surrounding the secondary reflector (which functions also as a cooled aperture stop), and the secondary reflector itself, then absorbs as much heat as possible from the structural box to reduce the radiative loading on colder stages, before returning in counterflow to the circulation compressor. This circulated-gas J-T loop would use $^3$He as the working fluid, whereas the MIRI cooler used $^4$He; the thermodynamic properties of $^3$He provide superior performance at the lower temperature. The performance attained at 4.5 K using $^3$He in this system is similar to that obtained at 6 K using $^4$He, so heat lift capability similar to that of MIRI can be anticipated with minimal changes. Optimizations would include redesigned gas heat exchangers for the $^3$He loop to best utilize the $^3$He gas properties. The gas distribution system would differ from MIRI because there is no on-orbit deployment required, but would have more points of heat exchange.

As noted, the cryocoolers delivered for MIRI could meet PICO requirements with minimal changes. Both NGAS (manufacturer of the MIRI coolers) and Ball Aerospace have further developed this cryocooler architecture, both with higher-compression-ratio J-T circulators which would provide lower J-T temperature, and Ball with a larger circulator which could provide greater heat lift, should that be required. Future trades of CADR performance improvement with lower temperature cADR precooling vs the cost of that lower precooling temperature promise additional options as the development of these mechanical coolers proceeds.

The 4 K cooler compressor is located on the spinning part of the spacecraft bus, on the warm side of the V-grooves. Electric power rejected by the cooler mechanical components, and all power rejected by all cryocoolers electronics, is transferred to the spacecraft heat rejection system.

## ACKNOWLEDGEMENTS

Work on the PICO is supported by NASA through grant #NNX17AK52G to the University of Minnesota. The research described in this paper was partially carried out at the Jet Propulsion Laboratory, a part of the California Institute of

Technology, under a contract with the National Aeronautics and Space Administration. Gianfranco de Zotti acknowledges financial support from the ASI/University of Roma-Tor Vergata agreement n. 2016-24-H.0 for study activities of the Italian cosmology community. Jacques Delabrouille acknowledges financial support from PNCG for participating to the PICO study. The information provided about the PICO mission concept is pre-decisional and is provided for planning and discussion purposes only.